\documentstyle[12pt]{article}
\newcommand{\leqsim}{\,\raisebox{-0.6ex}{$\buildrel < \over \sim$}\,}
\newcommand{\geqsim}{\,\raisebox{-0.6ex}{$\buildrel > \over \sim$}\,}

\def\etal{{\it et al}\,}
\def\hc{{\rm h.c.}}

\def\P{{\rm Pl}}

\def\gev{\,{\rm GeV}}

\def\sec{\,{\rm sec}}

\def\be{\begin{equation}}
\def\ee{\end{equation}}
\def\ba{\begin{eqnarray}}
\def\ea{\end{eqnarray}}
\renewcommand{\baselinestretch}{1.3}
\begin{document}

\pagestyle{plain}
\vspace{4cm}

\begin{center}

{\bf  IN SEARCH OF THE NEXT NEXT-TO-MSSM
\footnote{Talk presented at ,
``Electroweak Interactions and Unified Theories'', 
XXXIst Rencontres de Moriond, March 16-23, 1996} 
 }\\
\vspace{2 ex} 
\renewcommand{\baselinestretch}{1.3}
{
S.~A.~Abel\\
Service de Physique Th\'eorique\\
Universit\'e Libre de Bruxelles\\
Boulevard de Triomphe\\
Bruxelles 1050, Belgium
}\\

\end{center} 
\vspace{2cm}

\renewcommand{\baselinestretch}{1.0}
\begin{abstract}
\noindent It is argued that the Next-to-MSSM (NMSSM) is unnatural from 
the point of view of cosmology and
fine tuning. In particular, such singlet extensions to the MSSM do not
provide a simple solution to the `$\mu$-problem'.  
Models with singlets can be
constructed using gauged-$R$ symmetry or target space duality. However
their superpotentials have terms in addition to those of the NMSSM.  
\end{abstract}

\newpage
\section{Introduction}

According to current theoretical prejudice, the
Minimal Supersymmetric Standard Model (MSSM) is the most likely candidate for
physics beyond the standard model~\cite{mssm}. One of its most
interesting features is an upper bound on the mass of the lightest
higgs boson of about $120 \gev$ or so. This comes about because of 
the rather restricted form of the superpotential; 
\be
\label{mssmsuperpot}
W_{MSSM}=h_u Q_L H_2 U_R^c +h_d Q_L H_1 D_R^c +h_e L H_1 E_R^c + \mu
H_1 H_2,
\ee
where $Q_L$, $L$, $H_1$, and $H_2$ are the standard doublet
superfields. Because of this, and also because of the fine-tuning
problem inherent in the (phenomenologically necessary) choice of 
$\mu\sim M_W$ (sometimes referred to as the $\mu$-problem~\cite{muprob,gm}), 
many authors instead consider a variant of this superpotential known 
variously as the  Minimally-Extended-MSSM or Next-to-MSSM 
(NMSSM)~\cite{nmssm};
\be
\label{nmssmsuperpot}
W_{NMSSM}=h_u Q_L H_2 U_R^c +h_d Q_L H_1 D_R^c +h_e L H_1 E_R^c + \lambda
N H_1 H_2 -\frac{k}{3} N^3,
\ee
where $N$ is an additional singlet superfield. In these models the
$\mu$-term is generated radiatively when the latter aquires a vacuum
expectation value on the breaking of electroweak symmetry. Since
$\langle n \rangle \sim M_W$, this was thought to be a potential
simple solution to the $\mu$-problem.

Because of difficulties with cosmology (specifically the appearance of
domain walls) this appears to be no longer the case~\cite{aw,us},
however in view of the less constrained higgs phenomenology, it is
still worth pursuing models with singlet extensions.  Here I wish to
argue that the most natural way to avoid cosmological problems in
models with extra singlets is to introduce $\mu$ terms {\em in
addition} to the terms in eq.(\ref{nmssmsuperpot}). I shall show that
this may be done by borrowing from another solution to the
$\mu$-problem, proposed by Giudice and Masiero. Thus, if one wishes to
consider singlet extensions of the MSSM, it is more natural to
consider the most general form of the
superpotential~\cite{moorehouse},
\be
\label{nnmssmsuperpot}
W_{N^2MSSM}=h_u Q_L H_2 U_R^c +h_d Q_L H_1 D_R^c +h_e L H_1 E_R^c +
\mu H_1 H_2 + \mu' N^2+\lambda
N H_1 H_2 -\frac{k}{3} N^3.
\ee

The second point I shall discuss, is the fact that the singlet
field, $N$, is a potential hazard to the gauge-hierarchy. This is
because singlet fields can give rise to divergent tadpole diagrams.
Clearly, at the level of global supersymmetry, there is no
gauge-symmetry one can give to $N$ which forbids such operators
(that is linear $N$ terms) appearing. In fact there is a large number
of potentially dangerous operators which must be set to zero by hand
unaided by any symmetry. At the level of supergravity however, two
suitable symmetries become available. These are gauged $R$-symmetry, and 
target-space duality symmetry. In both these cases it is possible to
construct models which have no cosmological problems, no fine-tuning 
problems and no arbitrarily forbidden operators. 

\section{Cosmological Problems in the NMSSM}

First let us discuss the cosmological problems facing the simple
NMSSM. In addition to the terms derived from the superpotential in 
eq.(\ref{nmssmsuperpot}), there are soft-supersymmetry breaking terms.
It is these which are responsible for the breaking of electroweak
symmetry, and in this case, terms of particular importance are the
trilinear scalar couplings ($A$-terms) which appear in the potential. 
Together with the supersymmetric contribution, they lead to a scalar
potential of the form   
\ba
V_{soft} &=& -\lambda A_\lambda n h_1 h_2 -\frac{k}{3} A_k n^3
-\lambda k h_1 h_2 n^{*2} + \hc +\ldots \nonumber \\
         &=& -2 \mbox{\large{(}} \lambda A_\lambda \langle |n|\rangle \langle
           |h^0_1|\rangle\langle 
|h^0_2|\rangle \cos (\theta_1 +\theta_2 +\theta_n )
-\frac{k}{3} A_k \langle |n|\rangle ^3 \cos (3 \theta_n) \nonumber\\
& &  -\lambda k \langle |h^0_1|\rangle \langle |h^0_2|\rangle \langle
|n|\rangle ^2 \cos (\theta_1 +\theta_2 -2 \theta_n) \mbox{\large{)}} +\ldots
\ea  
where small letters indicate scalar components of superfields, and 
where for convenience I have taken $\lambda$, $k$, $A_\lambda$ and
$A_k$ to be real. The dots above stand for terms which either do not
get a VEV on electroweak symmetry breaking, or are independent
of the phases of the higgs scalars ($\theta_1$, $\theta_2$ and
$\theta_n$). For suitable choices of the parameters (e.g. all
positive) the above potential is clearly minimised where all the
cosines are +1 which has three solutions, namely $\theta_i=0$ or $\pm
2\pi /3 $. Field configurations which interpolate between any two
minima are topologically stable and lead to domain walls. Solving for
these is a relatively straightforward matter, and this was done in
ref.\cite{us} where, not surprisingly (given that all the VEVs and
$A_\lambda$ and $A_k$ are $\sim M_W$), it was found that the walls
have mass per unit area
\be
\sigma \sim M_W^3 \sim 10^5 \mbox{ kg } cm^{-2}.
\ee
Such walls are a cosmological disaster since, for example,  
their density falls as $T^2$ whereas that of radiation falls as $T^4$ 
so they eventually dominate and cause power law inflation~\cite{us}.

There are a number of solutions which one could consider to rectify
this situation. One which I shall not discuss in much detail here is
to embed the discrete symmetry in a broken gauge
symmetry~\cite{ls}. In this case the degenerate vacua are connected by
a gauge transformation in the full theory~\cite{ls,me}. After the
electroweak phase transition, one expects a network of domain walls
bounded by cosmic strings to form and then collapse. This situation
was examined in ref.\cite{me}, where the conclusion was that rather
complicated cosmological scenarios are required in order to be able to
accommodate it. The most natural solution, which is to simply insist
that the $Z_3$ symmetry be explicitly broken, will be the main focus
in what follows. 

\section{Breaking $Z_3$}

In principle, the $Z_3$ symmetry need not be broken by very much in
order to solve the domain wall problem. This was pointed out by 
Zel'dovich {\em et al} albeit in a rather different context~\cite{zko}, and
for the case at hand, it turns out that even gravitationally
suppressed terms are sufficient to remove the walls before the onset
of primordial nuclesynthesis (at $t\sim 1\sec$)~\cite{us,cosmobounds}.
(The release of entropy from walls collapsing after this time, would
effect the primordial abundances.) For example if one adds a piece 
\be
W_\epsilon = \lambda' \frac{N^4}{M_\P}
\ee
to the NMSSM superpotential, one requires only that $\lambda'\geqsim
10^{-7}$ in order to satisfy the above constraint. This is because the
walls continually straighten under their own tension, with the typical
radius of curvature increasing as $t$. Eventually even this tiny
pressure comes to dominate over the tension.

In ref.\cite{us} however, it was pointed out that this 
solution cannot work for the NMSSM. This is because all the 
operators suppressed by one power of $M_\P$  which one can write down,
lead to a divergent two or three loop diagram of the form discussed in 
ref.\cite{destab}. Such diagrams lead to a term linear in $n$;
\be
\delta V\sim \frac{\lambda'}{16 \pi^2} (n+n^*) M_\P M_W^2.
\ee
This destabilises the Planck/weak hierarchy unless 
$\lambda' \leqsim 10^{-11}$, but such a small value is clearly in 
conflict with 
the previous constraint coming from nucleosynthesis. 
(By simple power counting one finds that this is the leading
divergence which can occur.) 
These divergences can be calculated in the framework of $N=1$
supergravity~\cite{sriv}, in which the model depends only of the 
K\"ahler function 
\be
{\cal G} = K(\Phi,\overline{\Phi} ) +\log |\tilde{W}(\Phi)|^2 .
\ee
The function $K = \overline{K}$ is responsible for the kinetic terms,
and the superpotential $\tilde{W}$, is a holomorphic function of the
superfields (generically denoted by $\Phi$ above). However, as we
shall see, the effective low energy superpotential $W$ may receive terms
from $K$ as well as $\tilde{W}$.     

So any suitable $Z_3$-breaking model must have $\mu$ or
$\mu'\neq 0$ in the effective low energy lagrangian. In addition, any 
solution which can achieve this must of course also ensure the 
absence of the $N^4/M_\P$ operator above. In fact a brief examination 
of the possible divergences, shows that the operators 
$N H_i H^{\dagger}_i$ and $ N N
N^{\dagger}$ must be forbidden in $K$, and the following operators must
be forbidden in $\tilde{W}$;
\vspace{0.5cm}
\begin{center}
\begin{tabular}{||l|r|r||}   \hline
  \mbox{Operator}     & \mbox{Loop-order of divergent diagram} \\ \hline\hline 
    $N^2$, $H_1 H_2$     & 1         \\ \hline 
    $N^4$, $N^2 H_1 H_2$     & 2         \\ \hline 
         $(H_1 H_2)^2  $     & 3         \\ \hline
 $N^2 (H_1 H_2)^3$, $N^4 (H_1 H_2)^2$,
$N^6 (H_1 H_2)$, $N^8$     & 5         \\ \hline
$N^2 (H_1 H_2)^4$, $N^4 (H_1 H_2)^3$, $N^6 (H_1 H_2)^2$,
$N^8 (H_1 H_2)$, $N^{10}$     & 6         \\ \hline
\end{tabular}
\end{center}
\vspace{0.5cm}
In particular, the presence of the $N^2$ and $H_1 H_2$ operators in
this list means that the $\mu $ or $\mu'$ terms must come from $K$. To
achieve this without destabilising the hierarchy, one can simply add 
$H_1 H_2$ and/or $N^2$ into $K$ as was first suggested in ref.\cite{gm}. 
This indeed generates the desired $\mu$ and/or $\mu'\sim M_W$ terms in the
effective low energy (global supersymmetry) theory. 

However, an explanation for the absence of all the operators
above, requires an additional symmetry, under which $K$ and
$\tilde{W}$ transform differently. Two obvious examples are duality
symmetry and gauged-$R$ symmetry~\cite{herbi}. 
To conclude I shall present an example of the latter~\cite{me}.

In this case $K$ has zero $R$-charge, but $\tilde{W}$ has $R$-charge 
2. This means that the standard renormalisable NMSSM higgs superpotential,
\be 
\label{rwhiggs}
W_{\rm higgs}=\lambda N H_{1}H_{2}-
\frac{k}{3}N^3,
\ee
has the correct $R$-charge if $R(N)= 2/3$ and $R(H_1)+R(H_2) =4/3 $. 
So consider the K\"ahler function   
\be 
\label{quad}
{\cal G} = z^i z^\dagger_i + \Phi \Phi^\dagger 
+ \Phi ' \Phi ^{'\dagger}  
+ \left( \frac{\alpha}{M^2_\P}\Phi^\dagger\Phi'H_1 H_2 + 
 \frac{\alpha '}{M^2_\P}\Phi\Phi^{' \dagger}N^2 +\hc \right) 
+ \log |h(z ) + g(\Phi , \Phi' )|^2 ,
\ee
where $h(z)$ is the superpotential involving just visible sector fields 
and $\Phi$, $\Phi'$ here represent hidden sector fields with superpotential 
$g(\Phi , \Phi')$ (they may represent arbitrary functions of hidden 
sector fields in what follows). Both $\Phi $ and $\Phi^{'\dagger}$ appear 
here in order to prevent unwanted couplings being allowed in the 
superpotential which must be a holomorphic function of superfields.

The invariance of $K$ requires that $R(\Phi
)+R(\Phi^{'\dagger}) =4/3 $. If, for example, one chooses the
$R$-charges to be $R(\Phi )= 16/3 $, $R(\Phi ')= 4$, {\em all} of the
dangerous operators are forbidden~\cite{me}. The form of the low
energy superpotential is then that in eq.(\ref{nnmssmsuperpot}).
This, more general singlet extension of the MSSM, therefore appears 
to be a much more natural choice from the point of view of cosmology 
and fine-tuning. 

\newpage
\renewcommand{\baselinestretch}{1.0}

\end{document}